# A DDoS-Aware IDS Model Based on Danger Theory and Mobile Agents

## (Full version)[*]

Mahdi Zamani, Mahnush Movahedi
Department of Computer Science
University of New Mexico
Albuquerque, NM, USA
{zamani,movahedi}@cs.unm.edu

Mohammad Ebadzadeh, Hossein Pedram
Department of Computer Engineering
Amirkabir University of Technology
Tehran, Iran
{ebadzadeh,pedram}@ceit.aut.ac.ir

*Abstract*— We propose an artificial immune model for intrusion detection in distributed systems based on a relatively recent theory in immunology called Danger theory. Based on Danger theory, immune response in natural systems is a result of sensing corruption as well as sensing unknown substances. In contrast, traditional self-nonself discrimination theory states that immune response is only initiated by sensing nonself (unknown) patterns. Danger theory solves many problems that could only be partially explained by the traditional model. Although the traditional model is simpler, such problems result in high false positive rates in immune-inspired intrusion detection systems. We believe using danger theory in a multi-agent environment that computationally emulates the behavior of natural immune systems is effective in reducing false positive rates. We first describe a simplified scenario of immune response in natural systems based on danger theory and then, convert it to a computational model as a network protocol. In our protocol, we define several immune signals and model cell signaling via message passing between agents that emulate cells. Most messages include application-specific patterns that must be meaningfully extracted from various system properties. We show how to model these messages in practice by performing a case study on the problem of detecting distributed denial-of-service attacks in wireless sensor networks. We conduct a set of systematic experiments to find a set of performance metrics that can accurately distinguish malicious patterns. The results indicate that the system can be efficiently used to detect malicious patterns with a high level of accuracy.

## I. Introduction

Data and tasks in large-scale distributed systems are often enormously large that requires us to break them into tolerable pieces and placing them on various nodes. On the other hand, fragmentation and distribution of data and tasks over the network produces excessive overhead leading to reduction in performance and unavailability of data and tasks. This leads us to compromise between locality and distribution and maximize the locality to suppress intemperance in fragmentation [2].

Natural immune system is a distributed system that greatly balances locality and distribution. It consists of a vast number of cells performing their tasks concurrently and interacting with each other in various ways. The localized inter-cell interactions result in expression of complicated behaviors across the system that provide scalability and robustness. This fact has played an important role in motivating computer researchers in the last two decades to develop efficient distributed systems based on the structure and behavior of natural immune systems. *Artificial immune systems (AIS)* are computationally intelligent systems based on the behavior of the natural immune system. In other words, AIS are adaptive systems inspired by theoretical immunology and observed immune functions, principles and models, which are applied to complex problem domains. The AIS algorithms and models typically exploit the immune system's characteristics such as learning, memory and ubiquitousness to solve a complicated problem.

Intrusion detection is the process of dynamically monitoring events occurring in a computer system or network, analyzing them for signs of possible incidents and often interdicting the unauthorized access [21]. This is typically accomplished by automatically collecting information from a variety of systems and network sources, and then analyzing the information for possible security problems. Traditional intrusion detection and prevention techniques, like firewalls, access control mechanisms, and cryptography, have several limitations in fully protecting networks and systems from increasingly sophisticated attacks like denial of service. Moreover, most systems built based on such techniques suffer from high false positive and false negative detection rates and the lack of continuously adapting to changing malicious behaviors. An *intrusion detection system (IDS)* generally has to deal with problems such as large network traffic volumes, highly uneven data distribution, the difficulty to realize decision boundaries between normal and abnormal behavior, and a requirement for continuous adaptation to a constantly changing environment. In general, the challenge is to efficiently capture and classify various behaviors in a computer network.

*A. Our Contribution*

We design a general-purpose AIS inspired by several well-known immunological models that can be used for solving the intrusion detection problem. The main natural model we use is called *danger theory*, a relatively recent and strong model that explains many questions, which could, at best, be partially answered by the traditional *self-nonself discrimination theory*. In the simplest form, danger theory suggests that a natural immune response is the result of sensing danger in the system rather than detecting foreign symptoms of the cause, which is the basis of the self-nonself discrimination theory.

Although our model can be applied to any classification problem, several details remain to be defined by the specific application. We will show that one requires modeling several *immune signals* for knowledge representation as the *patterns* used in classification of data. Modeling immune signals is a delicate task and often requires several experiments to extract meaningful features of data.

In our previous work [24], we proposed an IDS model based on danger theory that is resilient to denial of service attacks in wireless sensor networks. We conducted pure optimization tests and applied experiments for detecting misbehaviors in wireless sensor networks. Wallenta *et al.* [22] and [16] proposed a similar work using the *dendritic cell algorithm (DCA)* for wireless sensor networks inspired by the Danger Theory. They argue that a general vulnerability in sensor network protocols is due to an interest cache with limited size. Although they analyzed the problem in detail and proposed an effective method accompanied by experiments, their model was only aimed at detecting attacks in wireless sensor networks. Our work is, however, different in many aspects. First, our model can be used for designing IDS for any distributed system. Second, our model is inspired by a more comprehensive immunity scenario described by the danger theory, not only the dendritic cells behavior. In our model, we have incorporated the behaviors of lymphocytes and dendritic cells in several tissues, e.g., clonal selection in lymph nodes and negative selection in thymuses. Second, we model communications that happen among various immune cells naturally performed via cell signaling. We believe these communications result in very sophisticated cooperation among the cells and are essential in detection of malicious behaviors.

In our protocol, we model immune signals as messages that can be exchanged between protocol agents. We show how to define these messages in practice by performing a set of statistical experiments on a wireless sensor network simulation. We conduct our experiments using *fractional factorial design* [12], a well-known performance evaluation technique used to minimize the number of experiments required to extract meaningful features.

*B. Paper Organization*

Section II covers the required background on immunology. Section III is a review of important related work. In section IV, we precisely define the assumptions required for our protocol to work. In section V, we propose our immune-inspired protocol. In this section, we convert the immune system scenario of section II into a computational model.

## II. NATURAL IMMUNE SYSTEMS

In this section, we describe the preliminaries required for understanding our model. Readers familiar with natural immune systems can skip this section.

*Natural immune system* consists of molecules, cells, and tissues that establish body's resistance to infection and inhibitory of decay agents in a complex fashion. *Lymphocytes* are white blood cells that play major roles in the immune system. They consist of two groups of small cells called *T-cells* and *B-cells* that form the *adaptive immune system*, the highly specialized immunity of vertebrates. Lymphocytes have receptors on their surface that are molecular structures (patterns) capable of binding to surface structures of other cells and *antigens*. Antigens are any molecular structure that can bind to lymphocyte receptors. An antigen that pertains to the body is called *self-antigen* [1]. Surface patterns of cell may become specific to antigenic patterns through selection processes. Once a sufficient number of an antigen is detected by the immune system, B-cells and a group of T-cells called *killer T-cells* ($T_K$) start the process of destroying the antigens. B-cells generate antigen-specific antibodies and release them in blood. Antibodies circulate the body and destroy pathogens they match with. In a similar process, killer T-cells adapt to pathogenic patterns and destroy the infected cells [19].

T-cells and B-cells are formed and matured in *thymus* and *bone marrow* respectively. Two important theories of immunology explain how lymphocytes mature and proliferate themselves. Based on the *negative selection theory*, immature cells with randomly generated molecular patterns are exposed to body cells. If they match the cells with a high affinity, they are condemned to death. After several attempts, a set of self-tolerated mature detector cells are generated. Based on the *clonal selection theory*, if the receptors of a lymphocyte match the molecular patterns on the surface of an antigen, the lymphocyte is activated to proliferate itself rapidly giving rise to selection of certain type of lymphocyte that is specific to the invading antigen [1].

Once the negative selection process is completed, mature lymphocytes migrate to *lymph nodes*; a special type of immune organs distributed across the body, where external antigens activate lymphocytes and the adaptive immune response is stimulated. Lymph nodes are connected to each other through a vascular system of *lymphatic vessels*. The major role of lymphatic vessels is transporting immune cells to and from lymph nodes.

In contrast to adaptive immune system, *innate immune system* is inherited from invertebrates and is basic and nonspecific. Innate immune system provides powerful and vital defense against infectious materials and plays a fundamental role in triggering adaptive immune response. On the other hand, adaptive immune system uses many of innate immune system mechanisms to remove detected infections. Generally, adaptive immune systems provide a slower but more precise defense to pathogens. *Dendritic cells (DCs)*, *macrophages,* and B-cells are innate immune cells that can engulf unknown pathogens and break them down into simple antigenic pieces called *peptides*. Due to their ability in presenting antigenic patterns to lymphocytes, these cells are known as *antigen-presenting cells (APCs)*. They also have special receptors on their surface, called *pattern recognition receptors (PRRs)* that play an important role in triggering innate immune response and consequently, adaptive immunity [19].

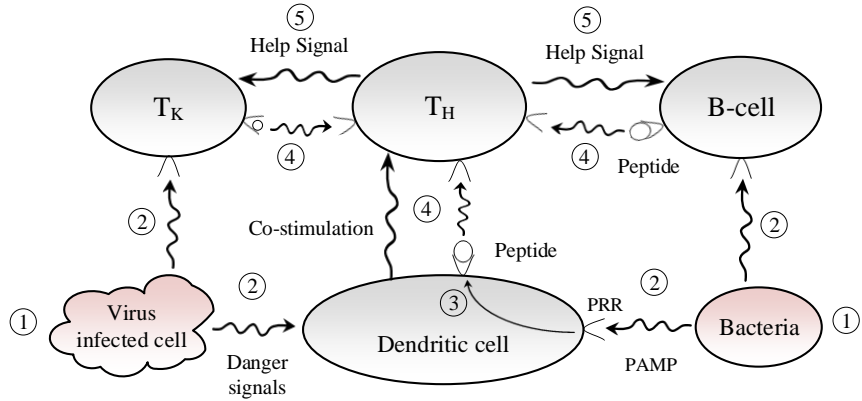

Figure 1. Immune response based on danger theory. Activated by danger and pathogenic signals, a dendritic cell migrates to a lymph node, where T-cells and B-cells exist. The dendritic cells, then, present pathogenic patterns to the lymphocytes and secrete co-stimulation signals to stimulates helper T-cells ($T_H$), which trigger immune response via killer T-cells ($T_K$) and B-cells to destroy infected cells.

Immune cells can communicate with each other in various ways. They may have either direct surface contacts or indirect chemical secretion contacts [9]. The former occurs when two cells collide and are capable of stimulating each other. In the latter way, a secreted chemical affects the cells in the near proximity of the generator, which have receptors sensitive to it. Secretive immune signals are called *cytokines* and encompass a large family of signals with different meanings [1].

B-cells and killer T-cells are only activated when an additional signal called *help signal* is generated near them by a group of T-cells called *helper T-cell* or $T_H$[*]. It is also discovered that in order to generate the help signal, $T_H$ requires a signal called *co-stimulation* generated by APCs. An APC is initially inactive and is activated by *pathogen-associated molecular patterns (PAMP)* that are sensed by its PRR.

For many years, immunologists used the self-nonself discrimination model to explain the reaction of immune system against pathogenic invasions. Based on the model, the immune system should learn to distinguish foreign materials from its own. The model is, unfortunately, unable to explain several observations. For example, it cannot explain why a self-cell remains self even though it changes throughout its life or why our bodies do not start any immune response against the foods we eat although they are actually foreign materials [18]. With an attempt to explain these, Matzinger [17] proposed a new model called *danger theory* where immune response is initiated as a result of sensing *danger* besides strange antigenic patterns. In other words, the simultaneity of foreign antigens and danger signals is crucial for initiation of immune response. Matzinger explains that any sign of abnormal cell death that is released around a cell can be regarded as danger signal. Danger, co-stimulation and help signals are generated sequentially and upon lack of any, the next signal and consequently the immune response would not start. Figure 1 shows the scenario of immune response based on danger theory. Five phases shown in the figure indicate the sequence of events:

1. In a local tissue[†], bacteria enter the body and viruses infect body cells[‡] and change their normal behavior.

2. Infected body cells release substances such as cell internal contents that act as danger signals. These substances stimulate dendritic cells that are present in the local tissue. Dendritic cells detect PAMPs on the surface of bacteria using their PRRs and engulf the corresponding antigens.

---

[*] Similar to killer T-cells, helper T-cells are also a heterogeneous group of T-cells.

[†] By tissue, we mean any part of the body that consists of an aggregate of cells having a similar structure and function. By local tissue we mean any tissue that is vulnerable to infections by bacteria or viruses.

[‡] Bacteria are unicellular living organisms while viruses are non-living organisms that are made of groups of RNAs or DNAs.

3. Dendritic cells analyze the antigens and break them up into peptides. They present the peptides on surface and migrate to lymph nodes, where lymphocytes exist.
4. Presented antigens stimulate matching T-cells. Meanwhile, dendritic cells express co-stimulatory molecules on their surface in proportion to the amount of danger and PAMP signals.
5. T-cells have molecules on their surface that can be stimulated by the co-stimulation molecules.
6. Activated T-cells proliferate and differentiate via clonal selection to precisely match the triggering antigenic patterns. Helper T-cells ($T_H$) release help signals to activate killer T-cells ($T_K$) and B-cells, which have already been stimulated by similar pathogens.

### III. RELATED WORK

Farmer *et al.* [5] proposed the first immune-inspired model. Inspired by the immune network theory they represented the dynamic behavior of the immune system by a set of differential equations and showed that they can be efficiently applied to learning problems. Forrest et al. [6] proposed the first immunological approach to computer security problems. Their work was fundamentally based on the self-nonself discrimination theory of immunology, which is based on the negative selection theory. Their approach was anomaly-based and they applied a simple computer model of negative selection as a change detector to the virus detection application. Despite many advantages of their scheme, it has some computational difficulties especially an exponential cost for detector generation.

Kephart proposed one of the earliest artificial immune systems for intrusion detection, which was to detect computer viruses and worms [14]. His method was a combination of signature and anomaly-based IDS with decoy programs that produced a set of antibodies to detect previously known viruses and some integrity monitors to detect changes to programs and data using checksums. Hofmeyr and Forrest [11] proposed the first distributed AIS applicable to various computer security problems. Their model is specialized to detect intrusions in local area networks. Each TCP connection is modeled by a triple, which encodes address of sender, address of receiver and port number of the receiver. Detectors are generated randomly through negative selection. In addition, they used a second signal called co-stimulation to confirm the anomaly that was detected via negative selection. In this system, a human is required to generate this signal manually in order to reduce false alarms (autoimmunity) of the system.

Kim and Bentley [15], assess the similarities between network-based IDSs and the human immune systems. They describe seven requirements of network IDSs: robustness, configurability, extendibility, scalability, adaptability, global analysis, and efficiency. They explain that the human immune system is distributed through immune networks and it generates unique antibody sets to provide the first four requirements. It is self-organized through gene library evolution, negative selection, and clonal. Finally, it is lightweight through approximate binding, memory cells, and gene expression to increase efficiency.

The first multi-agent IDS based on AIS was first proposed by Dasgupta [3] and then was continued in many other works like [10] and [23]. Dasgupta [3] defines three types of agents: *monitoring agents* that roam around the network and monitor various parameters simultaneously at multiple levels (user to packet level), *communicator agents* that are used to play the role of immune signals and *decision/action agents* to make decisions based on collected local warning signals. Harmer *et al.* [10], defined an agent-based architecture that maps required computer security elements to the immunological domain. Several matching rules for measuring affinity between detector strings and unknown strings are presented and analyzed. Like previous models, nonself strings are generated through negative selection. To overcome the problem of imperfect detector strings, helper agents ask for co-stimulation as a second confirmation signal, which should be generated manually by an administrator to reduce false-positive errors. One of the disadvantages of this model is that the main immune scenario does not perfectly match the natural one. Based on biological findings, antibodies are released from a differentiated type of activated B-cells after the clonal selection process. B-cells are activated by helper T-cells that are previously activated via co-stimulation. Greensmith et al. [7] and [8] suggested a novel anomaly detection algorithm inspired by functionalities of dendritic cells. They abstracted dendritic cells as signal processors that convert three input signals received by the cell to one output signal. The signals are based on the discoveries about biological concentration levels.

In 2005, Sarafijanovic and Boudec [20], developed an AIS for detection of misbehaving nodes in mobile ad-hoc networks. In these networks, nodes can be used as both hosts and routers and misbehaving node is the one that disobeys routing protocol. Misbehaving nodes participate in the routing process and drop routing control packets with a determinable probability. To discriminate between well-behaving and misbehaving nodes, a group of events related to the sent and received packets in a discrete time interval are labeled and collected in a sequence to form the antigens. Several routing protocol events in a discrete time interval are labeled and collected in a sequence to form the antigens. Mature detectors produced via negative selection monitor the network traffic. Once a match is occurred, the detector goes into a clonal selection algorithm and proliferates itself quickly. Danger signals are not used in this model; however, the authors recommend them to fortify the model. They suggest defining danger signals as network performance indicators such as packet loss and delay.

## IV. MODEL

We assume the distributed system is composed of several *nodes* that are connected via a synchronous network with all-to-all communication[*]. We also assume that at most a constant fraction of nodes is controlled by an adversary who uses the nodes to attack the network. We say that the nodes controlled by the adversary are *bad* and that the remaining nodes are *good* meaning that they strictly follow our protocol. For simplicity, we assume that the adversary cannot attack our protocol in any way but it can attack any other part of the system. We also assume that the adversary is *static* meaning that it must select the set of bad parties at the start of the protocol.

## V. OUR PROTOCOL

In this section, we describe a multi-agent protocol for intrusion detection in distributed systems based by the immune system scenario described in section II. We define two types of *agents* denoted by T and D corresponding to T-cell and Dendritic cell respectively. For simplicity in this paper, we only consider the problem of passive intrusion detection and we leave possible reactive strategies as a future work. Thus, we do not model the behaviors of B-cells and killer T-cells as they provide reactive immunity. We also define four types of nodes denoted by TM, BM, LN, and LT that correspond to the four types of tissues involved in the immune scenario respectively: thymus, bone marrow, lymph node, and local tissue. In a setup phase, three nodes are elected to play the roles of TM, BM, and LN. The rest of the nodes play the role of LT. At each round of the protocol, each node possesses a set of agents. Each agent is associated with a sequence of bits denoted by MP[†] that is generated uniformly at randomly when the agent is first created in BM. T agents born at a BM are initially immature and are matured in TM via the negative selection algorithm [6]. LN is the place where activated D agents gather to activate T agents. To improve performance and availability, several LN nodes can be created and distributed across the system.

### A. Modeling Cell Signaling

In order to model cell signaling in our protocol, we let each agent to probabilistically communicate with other agents that are in the same node it belongs. In other words, a message sent by an agent is delivered to a recipient with probability $p$ that is a parameter of our protocol. To model the propagation of natural immune signals, each agent chooses a set of $t$ agents from the node it belongs uniformly at random, where $t$ is a parameter of our protocol. The agent, then, sends its message (signal) to each of the $t$ agents with probability $p$. We simulate both types of cell signaling (i.e., secretion and direct contact) with the same method described above.

### B. Modeling Immune Signals

We model immune signals with messages that can be transmitted between agents. For simplicity, we only model three types of signals involved in the immune scenario shown in Fig 1: PAMP, danger, and co-stimulation. Recall that PAMP is a pattern associated with the molecules on a cell's surface. We model such a pattern with a multi-dimensional feature vector that we call *Antigenic Molecular Pattern (AMP)*. Both self and pathogenic patterns are modeled via AMP. AMPs must be extracted from application-specific events occurring in the system and made self-tolerant via the negative selection algorithm of [6] in TM or BM. This implies that a set of normal AMPs are necessary to train T agents.

We model a danger signal via a multi-dimensional feature vector that we call *Danger Molecular Pattern (DMP)*. Although DMP is a type of signal, it is not directly sent by any agent and it is rather *detected* by D agents. We explain this more when we model D agents later in this section. When modeling AMP and DMP in practice, it is crucial to remember the important difference between them: an AMP must reflect a sign of known malicious behavior but DMP must indicate symptoms of corruption that are not necessarily correlated with malicious patterns (i.e., AMPs). This plays an important role in reducing false positive rate of our protocol.

Recall that in natural immune systems, co-stimulation is a signal generated by dendritic cells to trigger $T_H$ cells and is required in addition to antigen-specific PAMP signals to activate $T_H$. Unlike PAMPs, co-stimulation signals do not carry any antigenic patterns and consist of only one type of protein[‡]. Hence, we model the co-stimulation signal with a message that only contains one field indicating the concentration (power) of co-stimulatory molecules[§].

### C. Modeling Dendritic Cells

For simplicity, we only model dendritic cells as the only type APC in our protocol. We model dendritic cells via D agent. Once a D agent receives a message, it analyzes the associated AMP using an application-specific multi-dimensional classifier to determine whether the AMP is pathogenic or not.

In addition to AMP, D agents can detect and analyze DMPs. Each D agent that is in an LT, always scans certain properties of the system to extract DMPs and detect corruption to the system. This means that D agent requires a

---

[*] Direct physical all-to-all links are not necessary. An overlay network providing all-to-all communication suffices.

[†] MP recalls molecular pattern. In immunology, it is a sequence of molecules on the surface of immune cells that bind to antigens.

[‡] APCs express a type of protein (CD86) on their surface as the co-stimulation signal.

[§] Concentration of co-stimulatory molecules is correlated with the concentration of PAMPs detected by the dendritic cell [7].

*DMP classifier*, which is application-specific. Each D agent must detect certain amounts of AMP and DMP to be activated. Once activated, it migrates to the nearest LN and publishes co-stimulation messages to nearby agents inside the LN[*]. In order to model the co-stimulation concentration (defined above), the agent uses a distance metric to measure how much the detected AMP is close to the training vectors[†]. Greensmith et al. [7] model the correlation between co-stimulation concentration and PAMP affinity and the amount of danger signals the dendritic cell detects.

### D. Modeling Cell Activation

Each T agent has an *activation threshold*, which is the number of matching AMPs it receives. The agent then, proliferates rapidly by creating new T agents[‡]. The proliferation rate is a parameter of our protocol and can be tuned by experiment. The proliferation step amplifies the effect of T agents in immune response. During this stage, each T agent runs the clonal selection algorithm of [4] to diversify the AMPs of the generated agents. This is subsequently followed by the selection process that is mainly a converging process. As a result, the new agents are accurately adapted to similar instances of the attack.

### E. Modeling Cell Lifespan

In order to maintain diversity and suppress undesirable deviations in population of agents, our protocol considers a lifespan for each agent by incrementing a monotonically-increasing counter in each round of the protocol. Once the counter reaches a threshold that is a parameter of the system, it stops all of its scheduled tasks and simply deallocates all the associated memory. This simulates the natural programmed cell death process called *apoptosis*[§].

## VI. EVALUATION

In order to evaluate the proposed algorithm, we perform a case study to show various steps needed in modeling immune signals for the problem of detecting *distributed denial-of-service (DDoS)* attacks in *wireless sensor networks (WSN)*. A WSN is a network of *spatially-distributed* autonomous *sensors* (also called *nodes*) that monitor environmental conditions like temperature and pressure. Each sensor uses local radio connections with other sensors to route the captured data throughout the network to a specific location. Typically, a WSN consists of a large number of low-cost sensors. Each sensor has access to a very limited amount of energy thus; WSNs should use energy efficient communication protocols and have self-organizing properties to maintain ubiquitousness of the network.

A DDoS attack occurs when many nodes launch a coordinated flooding attack against one or more targets to prevent them from doing their normal tasks. We hope that studying this specific problem will help the reader easily apply our immune-inspired scheme for other applications.

### A. Directed Diffusion Routing

Intanagonwiwat *et al*. [13] proposed a data-centric routing paradigm called directed diffusion to disseminate data in sensor networks. They define three types of sensor nodes: *data sinks* or *subscribers* are nodes that subscribe to a particular set of *interests* to receive specific data. *Data sources* or *publishers* are the nodes, which publish their data throughout the network. *Intermediary nodes* or *relays* forward the diffusion interests from sinks to sources. Sinks and sources may also relay diffusion information. Each node has a cache of distinct interests, called *interest cache*.

Sinks and sources send interest and exploratory data messages respectively. The initial flooding of interest and exploratory data together can build communication paths from sinks to sources and vice versa. Moreover, sinks can reinforce paths from neighbors by sending positive reinforcement message. Sinks may also negatively reinforce a neighbor path in case of better data delivery from other neighbors. Negative reinforcements suppress duplicate path and loops.

### B. Attack Scenario

Wallenta et al. [22] introduced a DDoS attack in diffusion-based WSN called *interest cache poisoning (ICP)*, which corrupts the routing process. In ICP, a malicious node floods the network by spurious interest messages. Since the interest cache in each node is finite, the correct interests are replaced by the spurious interests and hence, data messages corresponding to legitimate interests are dropped.

We perform several statistical experiments by simulating a wireless sensor network in ns-2 for both the normal case and the malicious case. In the malicious case, we assume a constant number of nodes are controlled by an adversary who corrupts the nodes by replacing their correct protocol procedures to induce ICP.

To model each message (immune signal) defined in section V, we extract a set of network metrics based on their sensitivity against several network parameters. In the following, we describe these metrics for DMP and AMP messages.

### C. Modeling Danger Molecular Patterns (DMP)

Danger patterns should indicate signs of corruption in the system. One good type of features that can show corruptions in computer networks effectively is *network performance metrics*. Many of these metrics change significantly after the network is attacked by an adversary. On the other hand, there are many of such metrics that makes it computationally hard to consider all of them in our DMP feature vectors. Moreover, many of those metrics do not change significantly by the ICP attacks. In order to pick the best set of metrics, we run a set of experiments on a group of nine candidate performance metrics defined in Table 1 against $k = 7$ network parameters shown in Table 2. We use the $2^{k-p}$ fractional factorial design of [12] to minimize the number of experiments needed, where $p$ is the reducer coefficient. We set $p = 4$ to make the number of experiments small (in this case 8). Since we repeat each experiment 5 times, the total number of runs is 40.

---

[*] A simple technique might be to ask D agents to record some information about the specific LT they migrate from in order to ban the malicious behavior.

[†] The distance metric depends on the classifier used in practice. Most statistical classifiers provide such a deviation measure to show the reliability of classification.

[‡] For simplicity, we only consider one type of T agent, which models a helper T cell.

[§] About 50-70 billion cells die each day in human adult via apoptosis.

Table 1. Candidate performance metrics (LB: Low Better, HB: High Better)

| Groups | Goal | Name | Description | |
|---|---|---|---|---|
| Speed | Responsiveness | Response time | Time between sending a data message from publisher and receiving it in sink | LB |
| | | One hop delay | Last hop delay average in one block | LB |
| | Productivity | Data throughput | Number of data messages received per block | HB |
| | | Interest throughput | Number of interest messages received per block | LB |
| | Utilization | Passed hops | Number of hops messages pass to arrive to destination sink nodes | LB |
| Reliability | Delay error | Delay probability | Probability that a message is delayed more than average response time along its way to destination sink node | LB |
| | Buffer lengthening | Long buffer probability | Probability that the send buffer length exceeds a specific threshold | LB |
| Availability | Loss error | Data drop rate | Number of messages dropped in a block due to non-matching subscriptions | LB |
| | Buffer overflow | Buffer drop rate | Number of messages dropped in a block due to buffer overflow | LB |

The main goal of the experiments was to find the metrics among those listed in Table 1, which satisfy the following conditions as much as possible:

1. They should change significantly when the value of $N_M$ changes, especially from $N_M = 0$ to $N_M > 0$ and vice versa and,

2. They should not change significantly when the value of $N_M$ is fixed and the value of other parameters listed in Table 2 change.

Table 2. Network parameters (factors)

| Factor | Description |
|---|---|
| $N_M$ | Number of malicious nodes |
| $N_P$ | Number of publish nodes |
| $I_P$ | Publish interval: Time interval between two publish messages sent by publish nodes. |
| $\mu_M$ | Mean of malicious distribution: Mean of the ratio of the number of malicious interests sent each time to the maximum send buffer length. |
| $N_R$ | Number of relay nodes |
| $N_S$ | Number of subscriber (sink) nodes |
| $\Delta\sigma_l$ | Initial phase difference: Time interval between two publish messages sent by publish nodes. |

In the fractional factorial design, for each metric a set of *significance levels* are calculated that each level is measured by the proportion of the total variation in the response that is explained by the factor. Table 3 shows the significant values for all metrics under test. These values were obtained by running the experiments described above and following the fractional factorial design of [12].

Table 3 shows that only three metrics have the two conditions explained above (i.e., change significantly with the ICP attack): *interest throughput*, *long-buffer probability*, and *one-hop delay*. Hence, we model DMP via a three-dimensional feature vector consisting of these three metrics. Based on the protocol, D agents are required to use a classifier to discriminate between safe and danger patterns, i.e. good and bad DMPs. To this end, we train a *Support Vector Machine (SVM)* using the a set of vectors (shown in Figure 2) obtained from 2800 ns-2 event blocks in which 1500 samples correspond to good samples (shown in blue) and 1300 samples correspond to malicious samples (shown in red). These samples are used to train a two-norm, soft-margin SVM with linear kernel. The figure also illustrates the classifier plane obtained by the SVM.

### D. Modeling Antigenic Molecular Patterns (AMP)

Unlike DMPs, AMPs are known malicious patterns and should be used in an initial decision. In other words, they should act as the signatures of misbehaving and well-behaving nodes. A suitable feature that has these properties is data throughput because it changes significantly in malicious nodes and does not change very much in good nodes.

### E. Results

Two major metrics of successful detection in the model are number of *activated* D *agents* and *number of activated* T *agents*. The expected behavior of these metrics is that they should increase monotonically by increasing the number of malicious sinks.

We performed a simulation of our protocol in both normal and malicious cases to evaluate the performance of the protocol in detecting malicious behaviors. Out of one hundred LT nodes in ten normal runs, only seven LT nodes on average contained activated D agents and only three LN nodes on average contained activated T agents (3% false positive). In the malicious case, 15 out of 100 LT nodes were malicious. We repeated the experiment ten times and on average, in 14 nodes, D agents were activated and all T agents of nearest LN nodes were activated (7% false negative). Figure 3 shows the number of activated D agents and T agents versus the number of malicious sink nodes. Both plots are monotonically increasing and have smooth exponential increase. One explanation to this fast increase is that each activated agent stimulates several other agents and hence, an exponential blowup is formed. This blowup can be relaxed by reducing each agents lifespan, which in return may increase the false negative rate.

Table 3. Metric significance values for malicious and non-malicious nodes

| Metric↓  Factor→ | Malicious Nodes | | | | | | | Non-malicious Nodes | | | | | | |
|---|---|---|---|---|---|---|---|---|---|---|---|---|---|---|
| | $N_M$ | $N_P$ | $I_P$ | $\mu_M$ | $N_R$ | $N_S$ | $\Delta\sigma_I$ | $N_M$ | $N_P$ | $I_P$ | $\mu_M$ | $N_R$ | $N_S$ | $\Delta\sigma_I$ |
| **Buffer Overflow** | 0.37 | 0.00 | 0.27 | 0.01 | 0.04 | 0.23 | 0.07 | 0.19 | 0.00 | 0.22 | 0.00 | 0.01 | 0.19 | 0.39 |
| **Data Drop** | 0.02 | 0.03 | 0.53 | 0.00 | 0.18 | 0.05 | 0.19 | 0.02 | 0.02 | 0.53 | 0.06 | 0.05 | 0.01 | 0.31 |
| **Data Throughput** | 0.72 | 0.03 | 0.07 | 0.15 | 0.01 | 0.00 | 0.02 | 0.29 | 0.06 | 0.14 | 0.02 | 0.07 | 0.02 | 0.40 |
| **Delay Prob.** | 0.23 | 0.00 | 0.17 | 0.08 | 0.07 | 0.31 | 0.12 | 0.17 | 0.01 | 0.20 | 0.02 | 0.08 | 0.37 | 0.15 |
| **Interest Throughput** | 0.64 | 0.01 | 0.02 | 0.01 | 0.03 | 0.20 | 0.09 | 0.63 | 0.02 | 0.03 | 0.02 | 0.04 | 0.18 | 0.08 |
| **Long Buffer Prob.** | 0.39 | 0.03 | 0.11 | 0.06 | 0.07 | 0.09 | 0.25 | 0.29 | 0.02 | 0.12 | 0.05 | 0.11 | 0.14 | 0.28 |
| **One Hop Delay** | 0.51 | 0.01 | 0.12 | 0.00 | 0.22 | 0.12 | 0.01 | 0.35 | 0.00 | 0.32 | 0.02 | 0.12 | 0.07 | 0.13 |
| **Passed Hops** | 0.05 | 0.26 | 0.26 | 0.03 | 0.03 | 0.32 | 0.05 | 0.09 | 0.22 | 0.22 | 0.08 | 0.08 | 0.22 | 0.09 |
| **Response Time** | 0.07 | 0.37 | 0.07 | 0.05 | 0.02 | 0.33 | 0.09 | 0.25 | 0.57 | 0.04 | 0.04 | 0.05 | 0.00 | 0.05 |

## VII. CONCLUSION AND FUTURE WORK

We proposed an immune-inspired protocol based on immunological danger theory for intrusion detection in distributed systems. Danger theory is able to solve many important problems in immunology that could only be partially explained by the traditional self-nonself discrimination theory. These problems often result in high false positive rates in immune-inspired IDS. We believe that using danger theory in a multi agent environment that simulates the natural immune system is effective in reducing false positive rates. In our protocol, we modeled several immune signals via messages that can be transmitted between the agents. Most messages contain feature vectors that must be meaningfully extracted from various system properties. Although we described a few rules of thumb to simplify this delicate task, one should consider application-specific strategies in order to extract the best features. On the other hand, we believe most distributed systems have several aspects in common and a detailed application of our protocol in one system can be very useful in applying the protocol to other systems. To this end, we performed a case study to apply our protocol to the problem of detecting intrusions in wireless sensor networks. The experiments indicate that the system can detect intrusions with small false positive/negative rates. As a future work, we are interested in conducting a set of rigorous experiments to find a group of performance metrics that can accurately distinguish adversarial patterns from normal ones and corruption (danger) patterns from safe patterns. Finally, we are interested to compare detection rates of our protocol to other immune-inspired schemes proposed in the literature.

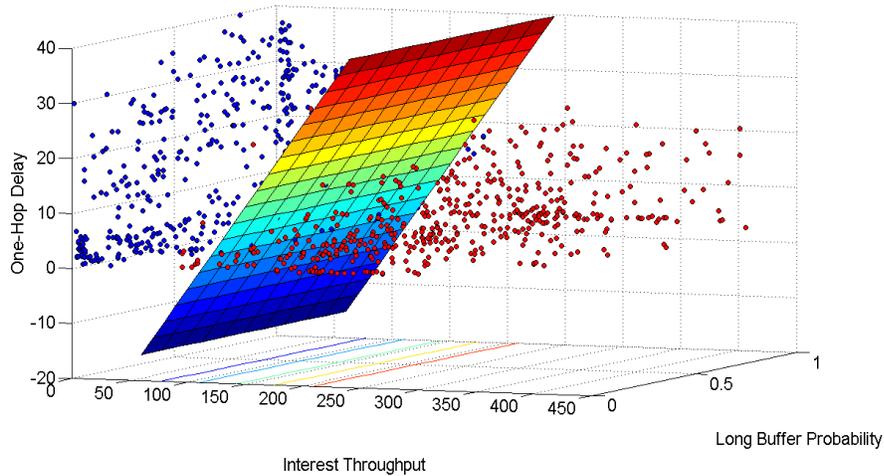

Figure 2. Modeling SDAMP signal: 1500 samples of event blocks obtained from healthy run (blue, left) and 1500 samples of event blocks obtained from a run with attack (red, right). The plane $-1.8x - 0.1y + 0.2z = 195$ resulted from training a SVM is used as the safe-danger classifier.

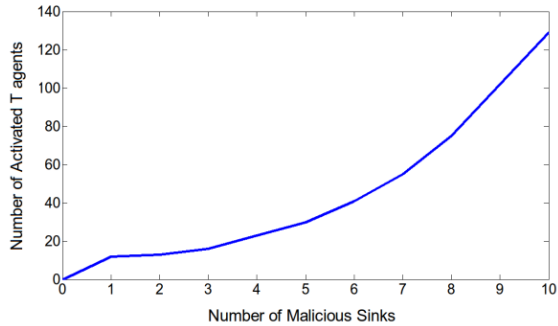 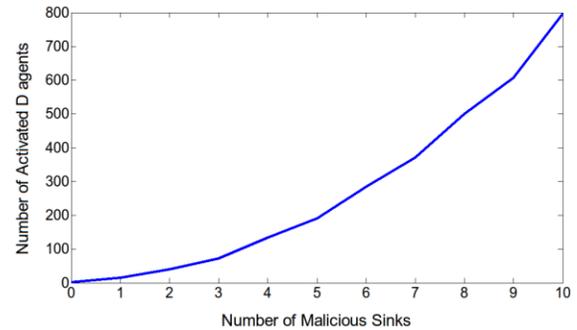

Figure 3. Number of activated D agents (right) and number of activated T agents (left) vs the number of malicious sinks

ACKNOWLEDGMENT

The authors would like to thank Mohammad Ebadzadeh and Hossein Pedram from Amirkabir University of Technology, Iran for their valuable contributions to the discussions and for their supportive comments.